\documentclass[twoside,journey]{IEEEtran}
\usepackage{makecell}
\usepackage{hyperref}
\usepackage{array}
\usepackage{graphicx,amssymb,amsmath}
\usepackage{multicol}
\usepackage[noadjust]{cite}
\usepackage{booktabs}
\usepackage{multirow}
\usepackage{tabularx}
\usepackage{enumitem}
\usepackage{setspace}
\usepackage{subfigure}
\usepackage{graphicx}
\usepackage{float}
\usepackage{url}
\usepackage[table,xcdraw]{xcolor}
\usepackage{stfloats}
\usepackage{amsthm,pifont}
\usepackage{flushend}
\usepackage{cases,subeqnarray}
\usepackage{bm,multirow,bigstrut}
\usepackage{amsmath, amsthm, amssymb}
\usepackage{textcomp}
\usepackage{latexsym,bm}
\usepackage{booktabs}
\usepackage{xcolor}
\usepackage{mathtools}
\usepackage{dsfont}
\usepackage{array}
\usepackage{booktabs}
\usepackage{makecell}
\usepackage{extarrows}
\usepackage{epsfig}
\usepackage{epsfig}
\usepackage{epstopdf}
\usepackage[noend]{algpseudocode}
\usepackage{algorithmicx,algorithm}
\usepackage{svg}

\theoremstyle{plain}

\theoremstyle{plain}

\usepackage{amsmath}

\IEEEoverridecommandlockouts

\DeclareUnicodeCharacter{3000}{~}

\begin{document}


\title{Generative AI-enabled Vehicular Networks: Fundamentals, Framework, and Case Study}
\author{Ruichen Zhang, Ke Xiong, Hongyang~Du,  Dusit~Niyato,~\IEEEmembership{Fellow,~IEEE}, \\ 
Jiawen~Kang, Xuemin~Shen,~\IEEEmembership{Fellow,~IEEE}, H. Vincent Poor, ~\IEEEmembership{Life Fellow,~IEEE}

\thanks{R.~Zhang, and K. Xiong are with the Engineering Research Center of Network Management Technology for High Speed Railway of Ministry of Education, School of Computer and Information Technology, Beijing Jiaotong University, Beijing 100044, China, with the Collaborative Innovation Center of Railway Traffic Safety, Beijing Jiaotong University, Beijing 100044, China, and also with the National Engineering Research Center of Advanced Network Technologies, Beijing Jiaotong University, Beijing 100044, China (e-mail: ruichen.zhang@bjtu.edu.cn, kxiong@bjtu.edu.cn). }
\thanks{H.~Du, and D. Niyato are with the School of Computer Science and Engineering, the Energy Research Institute @ NTU, Interdisciplinary Graduate Program, Nanyang Technological University, Singapore (e-mail: hongyang001@e.ntu.edu.sg, dniyato@ntu.edu.sg).}
\thanks{J. Kang is with the School of Automation, Guangdong University of Technology, China. (e-mail: kavinkang@gdut.edu.cn)}
\thanks{X. Shen is with the Department of Electrical and Computer Engineering, University of Waterloo, Canada (e-mail: sshen@uwaterloo.ca)}

\thanks{H. V. Poor is with the Department of Electrical and Computer Engineering, Princeton University, Princeton, NJ08544, USA (e-mail: poor@princeton.edu).}}
\maketitle

\begin{abstract}

Recognizing the tremendous improvements that the integration of generative AI can bring to intelligent transportation systems, this article explores the integration of generative AI technologies in vehicular networks, focusing on their potential applications and challenges. Generative AI, with its capabilities of generating realistic data and facilitating advanced decision-making processes, enhances various applications when combined with vehicular networks, such as navigation optimization, traffic prediction, data generation, and evaluation. Despite these promising applications, the integration of generative AI with vehicular networks faces several challenges, such as real-time data processing and decision-making, adapting to dynamic and unpredictable environments, as well as privacy and security concerns. To address these challenges, we propose a multi-modality semantic-aware framework to enhance the service quality of generative AI.  By leveraging multi-modal and semantic communication technologies, the framework enables the use of text and image data for creating multi-modal content, providing more reliable guidance to receiving vehicles and ultimately improving system usability and efficiency. To further improve the reliability and efficiency of information transmission and reconstruction within the framework, taking generative AI-enabled vehicle-to-vehicle (V2V) as a case study, a deep reinforcement learning (DRL)-based approach is proposed for resource allocation. Finally, we discuss potential research directions and anticipated advancements in the field of generative AI-enabled vehicular networks.

\end{abstract}


\IEEEpeerreviewmaketitle
\section{Introduction}
The rapid advances in the Internet of Things (IoT) and connected devices have spurred the development of the vehicular networks as a critical component of modern transportation systems. A vehicular networks facilitates seamless communication among vehicles, infrastructure, and other connected devices, with the potential to significantly enhance the efficiency, safety, and sustainability of transportation. By enabling real-time data exchange and collaborative decision-making, a vehicular networks can support a wide range of applications, such as traffic management, autonomous driving, and smart city planning \cite{8767077}. 

On the other hand, with the development of artificial intelligence (AI) technology, generative AI, as a branch of AI, has been receiving increasing attention. In particular, generative AI focuses on creating new content or data, such as images, texts, audios, videos, or even system designs, by learning patterns and structures from existing data and autonomously generating novel outputs. For example, ChatGPT\footnote{https://chat.openai.com/}, a generative AI language model developed by OpenAI, is capable of generating human-like text based on a given context. Also, some generative AI techniques such as generative adversarial networks (GANs), variational autoencoders (VAEs), and diffusion models can be utilized to generate realistic traffic simulation scenarios, create alternative routes for connected vehicles, or minimize energy consumption in electric vehicles. For example,  in \cite{Kim_2021_CVPR}, a DriveGAN was proposed to generate high-resolution and diverse simulations based on user-defined driving conditions, such as weather and positioning. Also,  in \cite{Yang_2020_CVPR},  a realistic sensor data synthesis framework was proposed for enhancing autonomous driving, where the data-driven camera generation approach not only produces visually high-quality data but also serves as training datasets that enhance the performance of AI algorithms.

The potential integration of generative AI into vehicular networks systems can lead to substantial advancements in intelligent transportation systems. By automating the content creation process and customizing it to users' personalized and proactive AI requirements, the effectiveness of vehicular networks-based services can be significantly improved \cite{mou2023t2iadapter}. For instance, the generative AI could automatically generate pertinent and timely information for drivers, such as real-time traffic updates, weather forecasts, and nearby amenities.  Furthermore, it could also be utilized to generate specialized driving data, such as accident simulations, to enhance real-world operational efficiency. Companies like Siemens Mobility Corp. leveraged AI techniques to improve traffic management by generating reliable real-time traffic predictions, facilitating better routing decisions, and minimizing congestion.\footnote{https://press.siemens.com/global/en/pressrelease/siemens-mobility-showcases-solutions-reduce-traffic-congestion} Similarly, Waymo Corp. released a VectorNet model to predict the behavior of traffic agents and their interactions with the surrounding environment in autonomous driving scenarios.\footnote{https://blog.waymo.com/2020/05/vectornet.html} Thus, generative AI can ultimately improve the efficiency and convenience of vehicular networks-based services by automating specific aspects of the content creation process and tailoring it to users' requirements.

\begin{table*}[htbp]
\caption{The advantages and disadvantages for different generative AI technologies}
\label{tab_PS}
\centering
\renewcommand{\arraystretch}{1.1} 
 \scalebox{0.70}{\begin{tabular}{|cll|}
\hline
\rowcolor[HTML]{ECF4FF} 
\multicolumn{3}{|c|}{\cellcolor[HTML]{ECF4FF}\textbf{ \large Data-based Generative AI technologies}} \\ \hline
\rowcolor[HTML]{ECF4FF} 
\multicolumn{1}{|c|}{\cellcolor[HTML]{ECF4FF}\textbf{Technologies}} &
  \multicolumn{1}{c|}{\cellcolor[HTML]{ECF4FF}\textbf{Advantages}} &
  \multicolumn{1}{c|}{\cellcolor[HTML]{ECF4FF}\textbf{Disadvantages}} \\ \hline
\rowcolor[HTML]{ECF4FF} 
\multicolumn{1}{|c|}{\cellcolor[HTML]{ECF4FF} Neural style transfer \cite{8732370}} &
  \multicolumn{1}{l|}{\cellcolor[HTML]{ECF4FF}\begin{tabular}[c]{@{}l@{}}-Ability to combine style and content from different images\\ -Real-time style transfer with optimized algorithms\\ -Applicable to various forms of media\end{tabular}} &
  \begin{tabular}[c]{@{}l@{}}-Dependence on high-quality style and content images\\ -Difficulty in preserving fine details\\ -Difficulty in controlling the level of stylization\end{tabular} \\ \hline
\rowcolor[HTML]{ECF4FF} 
\multicolumn{1}{|c|}{\cellcolor[HTML]{ECF4FF}Pixel CNN/Pixel RNN \cite{Kong_2018_CVPR}} &
  \multicolumn{1}{l|}{\cellcolor[HTML]{ECF4FF}\begin{tabular}[c]{@{}l@{}}-Autoregressive modeling of image pixels\\ -Explicitly captures local dependencies in data\\ -Effective in modeling discrete data\end{tabular}} &
  \begin{tabular}[c]{@{}l@{}}-Computationally expensive\\ -Difficulty in capturing long-range dependencies\\ -Sequential nature limits generation speed\end{tabular} \\ \hline
\rowcolor[HTML]{ECF4FF} 
\multicolumn{1}{|c|}{\cellcolor[HTML]{ECF4FF}Transformer-based model \cite{van2023chatgpt}} &
  \multicolumn{1}{l|}{\cellcolor[HTML]{ECF4FF} \begin{tabular}[c]{@{}l@{}}-Effective modeling of long-range dependencies\\ -Scalability with parallelization techniques\\ -Superior performance on various tasks\end{tabular}} &
  \begin{tabular}[c]{@{}l@{}}-Memory consumption in large-scale applications\\ -Complexity increases with sequence length\\ -Dependence on large training datasets\end{tabular} \\ \hline
\rowcolor[HTML]{ECF4FF} 
\multicolumn{1}{|c|}{\cellcolor[HTML]{ECF4FF}Example-based synthesis \cite{10.1145/2366145.2366154}} &
  \multicolumn{1}{l|}{\cellcolor[HTML]{ECF4FF}\begin{tabular}[c]{@{}l@{}}-Leveraging existing data for generation\\ -Ability to produce high-quality results\\ -Flexibility in incorporating domain-specific constraints\end{tabular}} &
  \begin{tabular}[c]{@{}l@{}}-Dependence on the quality and variety of input examples\\ -Difficulty in generalizing to unseen data\\ -Computational complexity\end{tabular} \\ \hline
\rowcolor[HTML]{FFF3E4} 
\multicolumn{3}{|c|}{\cellcolor[HTML]{FFF3E4}\textbf{\large Model-based Generative AI Technologies}} \\ \hline
\rowcolor[HTML]{FFF3E4} 
\multicolumn{1}{|c|}{\cellcolor[HTML]{FFF3E4}\textbf{Technologies}} &
  \multicolumn{1}{c|}{\cellcolor[HTML]{FFF3E4}\textbf{Advantages}} &
  \multicolumn{1}{c|}{\cellcolor[HTML]{FFF3E4}\textbf{Disadvantages}} \\ \hline
\rowcolor[HTML]{FFF3E4} 
\multicolumn{1}{|c|}{\cellcolor[HTML]{FFF3E4}Generative adversarial
network \cite{Kim_2021_CVPR}} &
  \multicolumn{1}{l|}{\cellcolor[HTML]{FFF3E4}\begin{tabular}[c]{@{}l@{}}-Exceptional quality of generated images\\ -Can incorporate conditions to control generated output \\ -Ability to address mode collapse with advanced techniques\end{tabular}} &
  \begin{tabular}[c]{@{}l@{}}-Training instability\\ -Difficulty in evaluating model quality\\ -Susceptibility to mode collapse\end{tabular} \\ \hline
\rowcolor[HTML]{FFF3E4} 
\multicolumn{1}{|c|}{\cellcolor[HTML]{FFF3E4}Variational Autoencoder \cite{pmlr-v80-jin18a}} &
  \multicolumn{1}{l|}{\cellcolor[HTML]{FFF3E4}\begin{tabular}[c]{@{}l@{}}-Effective representation of data in a probabilistic latent space\\ -Achieving disentangled representations\\ -Unified framework for both inference and generation\end{tabular}} &
  \begin{tabular}[c]{@{}l@{}}-Blurry image generation\\ -Limited expressive power of the prior distribution\\ -Difficulty in choosing likelihood functions\end{tabular} \\ \hline
\rowcolor[HTML]{FFF3E4} 
\multicolumn{1}{|c|}{\cellcolor[HTML]{FFF3E4}Autoregressive model \cite{du2023enabling}} &
  \multicolumn{1}{l|}{\cellcolor[HTML]{FFF3E4}\begin{tabular}[c]{@{}l@{}}-Direct modeling of temporal dependencies\\ -Scalability achieved through parallelization techniques\\ -Compatibility with both discrete and continuous data\end{tabular}} &
  \begin{tabular}[c]{@{}l@{}}-Sequential nature limits speed\\ -Complexity increases with sequence length\\ -Difficulty in modeling long-range dependencies\end{tabular} \\ \hline
\rowcolor[HTML]{FFF3E4} 
\multicolumn{1}{|c|}{\cellcolor[HTML]{FFF3E4}Diffusion-based model  \cite{mou2023t2iadapter}} &
  \multicolumn{1}{l|}{\cellcolor[HTML]{FFF3E4}\begin{tabular}[c]{@{}l@{}}-Generative process based on denoising score matching\\ -Flexible noise schedule selection\\ -Resistance to mode collapse and overfitting\end{tabular}} &
  \begin{tabular}[c]{@{}l@{}}-Slower generation process\\ -Complexity in training and selecting hyperparameters\end{tabular} \\ \hline
\end{tabular}}
\end{table*}

Despite the numerous advantages offered by generative AI-enabled vehicular networks, it encounters several challenges and limitations during implementation. The most prominent issue is limited communication bandwidth, which may lead to unreliable data transmission and communication security concerns within the vehicular networks ecosystem. Additionally, latency in communication between vehicles and infrastructure, coupled with the intricacy of data processing, poses significant barriers to the widespread adoption of vehicular networks.  Motivated by these, this article is an attempt to provide a forward-looking research for leveraging generative AI in vehicular networks systems. {\bf First}, we introduce different types of generative AI techniques and then describe the potential applications and challenges of these techniques in vehicular networks scenarios. \emph{To the best of the authors' knowledge, this is the first article that presents synergy between generative AI and vehicular technologies.} {\bf Second}, to address these challenges,  a novel generative AI-enabled vehicular networks framework that integrates a multi-modal architecture is proposed, which is capable of accommodating both text and image data, ultimately offering more reliable guidance to receiving vehicles. {\bf Third}, to further improve the reliability and efficiency of information transmission with the framework,  a deep reinforcement learning (DRL)-based approach is proposed to maximize the definite system quality of experience (QoE) within the constraints of the transmission power budget and the probability of successful transmission for each vehicle. Our approach jointly optimizes transmission communication resources and generative AI resources and simulation results validate the effectiveness of the proposed approach.

\section{Generative AI-enabled vehicular networks}
In this section, the various types of generative AI technologies is presented. Subsequently, we explore the key applications of these techniques in vehicular networks scenarios. Lastly, we discuss the prevailing challenges associated with implementing generative AI-enabled vehicular networks.

\subsection{Generative AI technologies}

As an emerging subfield of AI, generative AI technology focuses on autonomously generating new content or data, including images, text, audio, video, and even system designs. Similar to traditional AI technologies, such as discriminative AI, generative AI technologies can also be classified into two categories, i.e.,  the model-based technologies and the data-based technologies. Specifically, data-based generative AI employs learning algorithms to create novel content from existing data, while model-based generative AI leverages predefined models and simulations to generate new outputs by manipulating model parameters. These techniques excel in learning patterns and structures from existing data, producing outputs that closely resemble real-world samples, and enabling machines to mimic human creativity, imagination, and problem-solving abilities.

Developing a deep understanding of generative AI techniques suitable for specific applications in vehicular networks is crucial for optimizing performance and user experience. For instance, GANs can be utilized for data augmentation, enhancing the performance of AI models in tasks such as object detection or traffic prediction systems \cite{Kim_2021_CVPR}. VAEs are valuable for dimensionality reduction and feature extraction in sensor data, enabling more efficient processing and analysis \cite{pmlr-v80-jin18a}. Recognizing the strengths and limitations of each generative AI technique is essential for developers when implementing the most suitable strategy for specific applications within vehicular networks, ultimately improving overall system performance and user satisfaction. To provide a clear overview, the advantages and disadvantages of various Generative AI technologies are summarized in Table \ref{tab_PS}, which can better harness the potential of generative AI-enabled vehicular networks and related applications.

\begin{figure*}[t!]
\centerline{\includegraphics[width=\textwidth]{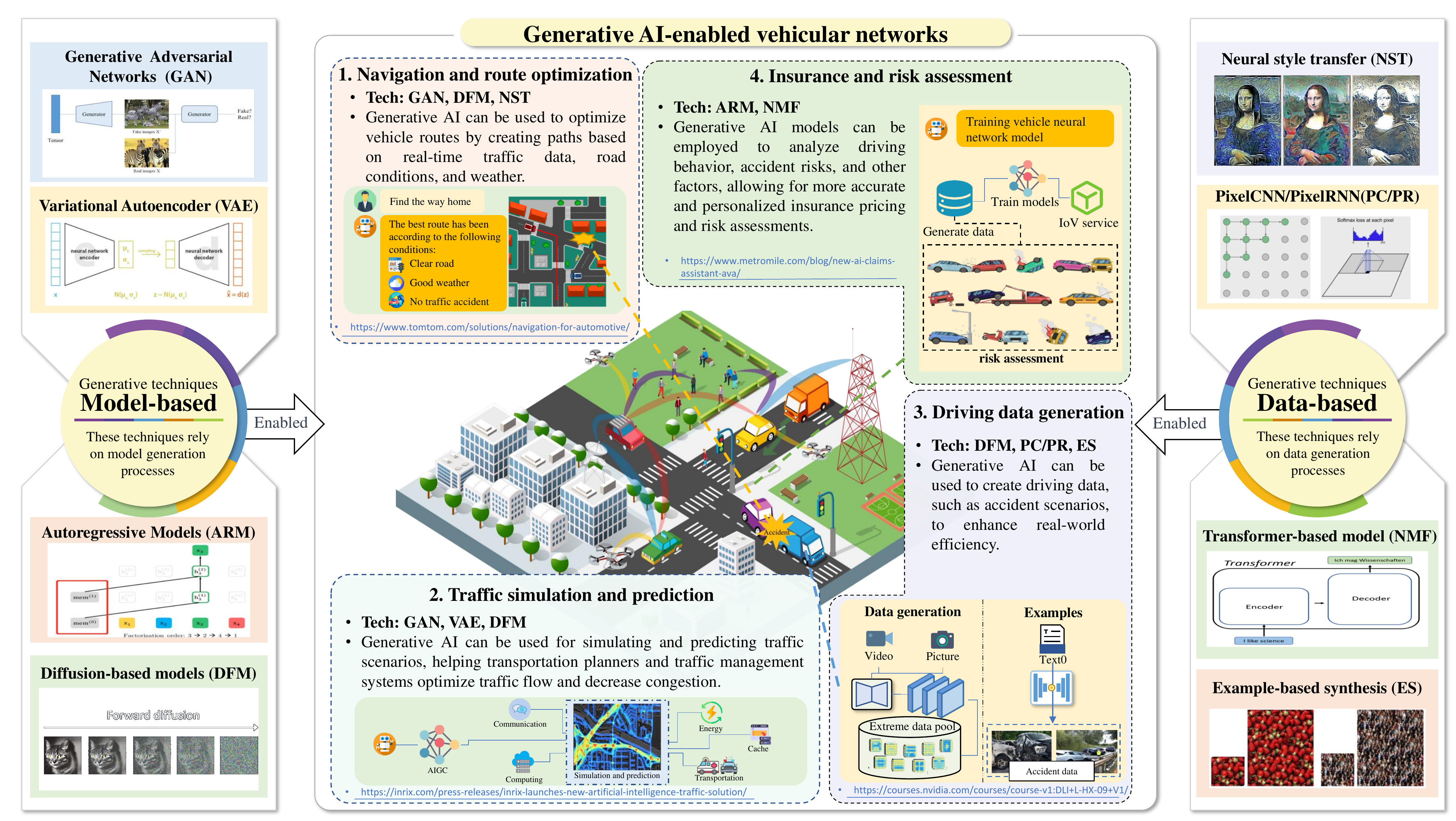}}
\caption{The schematic of generative AI-enabled vehicular networks. }
\label{Tutorial}
\end{figure*}
\subsection{Applications}

As shown in Fig. \ref{Tutorial}, by incorporating these advanced generative AI techniques, generative AI-enabled vehicular networks is capable of analyzing traffic patterns and providing drivers with real-time suggestions for alternative routes to circumvent congestion or accidents. For clarity, some major applications of generative AI-enabled vehicular networks are summarized as follows:

\begin{itemize}
\item {\bf Navigation and route optimization:} Generative AI can be used to optimize vehicle routes by creating paths based on real-time traffic data, road conditions, and weather. By effectively analyzing data and making quick decisions, generative AI can offer optimal routes for vehicles, improving overall driving efficiency and safety.  Furthermore, generative AI provides vehicles with real-time information about alternative routes and suggested driving speeds, helping them make informed decisions. In case of unexpected events like road closures, generative AI can quickly recalculate the best route for vehicles and adjust the path accordingly. For instance, Tomtom, a navigation system that uses generative AI,\footnote{https://www.tomtom.com/newsroom/behind-the-map/artificial-intelligence-map-making/} adjusts routes dynamically by using user-generated data and employing GANs to predict traffic patterns,  which allows drivers to avoid congested areas and reduce travel time. Thus, the potential of generative AI technologies for navigation and route optimization could lead to smarter and more efficient transportation systems for vehicular networks.

\item {\bf Traffic simulation and prediction:}  Generative AI can be used for simulating and predicting traffic scenarios, helping transportation planners and traffic management systems optimize traffic flow and decrease congestion. Specifically, generative AI models can examine historically traffic data to create reliable simulations of various traffic conditions, considering factors like time of day and road network configurations.  Additionally, generative AI can be applied to forecast future traffic patterns and demands, enabling proactive adjustments in traffic signal timings, dynamic rerouting, and effective resource allocation. For instance,  INRIX, a leading provider of real-time traffic and transportation data,\footnote{https://inrix.com/press-releases/inrix-launches-new-artificial-intelligence-traffic-solution/} adopts RNN/transformer-based generative models to predict traffic congestion levels in real-time, allowing city planners to make informed decisions about traffic management. By using generative AI technology in traffic simulation and prediction, transportation systems can become more responsive and adaptive to changing conditions, ultimately leading to improved traffic flow and reduced congestion.

\item {\bf Driving data generation:} Generative AI can be used to create driving data, such as accident scenarios, to enhance road safety. In particular, this driving data is employed to train generative machine learning models, improving their reliable in predicting and preventing accidents during real-world driving. For example, NVIDIA has developed a method called ``data augmentation through GANs" to generate synthetic driving situations, including rare or dangerous conditions, which can be used to train and refine autonomous vehicular systems.\footnote{https://courses.nvidia.com/courses/course-v1:DLI+L-HX-09+V1/} By generating a large amount of diverse and representative data, generative AI helps close the gap between simulated and real-world data, making machine learning models more robust and adaptable. Moreover, generative AI can address the challenge of data scarcity in vehicular networks, as collecting real-world driving data is usually costly and time-consuming. By leveraging generative AI technology in driving data generation, researchers and engineers can develop more efficient and safer vehicular systems, ultimately benefiting users and the transportation industry.

\item {\bf Insurance and risk assessment:} Generative AI models can be employed to analyze driving behavior, accident risks, and other factors, allowing for more reliable and personalized insurance pricing and risk assessments. These models can examine large amounts of historical driving data and generate synthesized instances that represent the underlying patterns and structures, which helps insurers better understand the factors contributing to accidents and other risks and adjust their pricing accordingly. For instance, Metromile, a pay-per-mile insurance provider, utilizes generative AI to analyze driver data and create personalized premiums based on individual driving habits and accident risk profiles.\footnote{https://www.metromile.com/blog/new-ai-claims-assistant-ava/} By harnessing generative AI in insurance and risk assessment, companies can develop more precise and equitable pricing models, ultimately fostering a more efficient and fair insurance market.

\end{itemize}

\subsection{Existing Challenges}

While the application of generative AI in vehicular networks has exhibited substantial potential, it also presents certain challenges that can be organized into three primary domains as follows.

\begin{itemize}
\item {\bf Real-time data processing and decision making:} Generative AI models in vehicular networks necessitate the ability to process and analyze vast amounts of data in real-time, consuming considerable bandwidth in the process. Especially for AI applications that involve high-resolution images or large data streams, both the upload and download processes require significant network resources to ensure low-latency services. For instance, high-definition maps used by autonomous vehicles, such as those provided by HERE Technologies,\footnote{https://www.here.com/platform/map-rendering} may consist of detailed data layers with file sizes ranging from several to tens of megabytes. Moreover, due to the dynamic nature of vehicular networks and the need for up-to-date information, vehicles may make multiple repeated requests to specific edge servers to obtain reliable data. 


\item {\bf Adapting to dynamic and unpredictable environments:} Vehicular networks are exposed to dynamic and often unpredictable environments, which necessitate generative AI models to rapidly adapt and respond to varying conditions. Specifically, the channel quality of vehicular networks can be severely impacted by the mobility of the vehicles themselves or changes in the surrounding environment, which affects the QoE of AI-enabled vehicular networks and further lead to decreased user satisfaction.


\item  {\bf Privacy and security:} Addressing data privacy and security challenges in generative AI models for vehicular networks is critical, as these systems deal with sensitive user information. The large-scale data collection and processing by connected vehicular  service providers pose concerns regarding personal information protection, as mandated by data protection laws. Specifically, vehicle-to-everything (V2X) communication systems, enabling data exchange between vehicles and connected devices, may inadvertently expose user information (e.g., license plate numbers, facial features, and vehicle details) to potential attackers, resulting in data breaches or privacy violations. Additionally, the presence of multiple parties in vehicular networks complicates interest alignment, with enhanced security potentially diminishing privacy. 

\end{itemize}

\begin{figure*}[t!]
\centerline{\includegraphics[width=\textwidth]{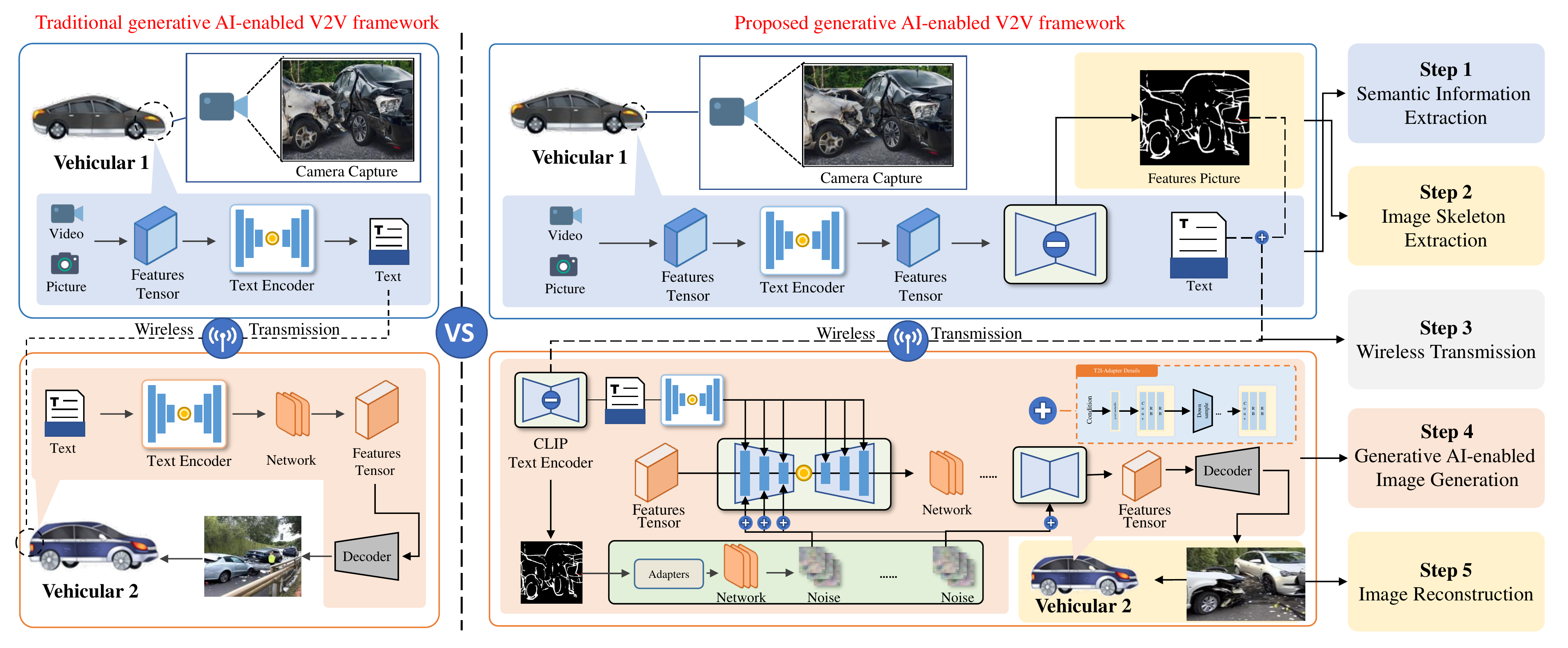}}
\caption{The framework of generative AI-enabled V2V framework. Part A: the traditional generative
AI-enabled V2V framework. Part B: Multi-modality Semantic-aware Framework for generative
AI-enabled vehicular networks.}
\label{Framework}
\end{figure*}


\section{Semantic-aware Framework for generative AI-enabled vehicular networks}

In this section, we propose a semantic-aware framework for generative AI-enabled vehicular networks to address current challenges and enhance road safety. 

\subsection{Conventional Semantic-aware Framework for generative AI-enabled vehicular networks}

Vehicle-to-vehicle (V2V) communication within vehicular networks has emerged as a promising approach to improve road safety. One potential involves using multiple cameras in vehicles to capture real-time images of road conditions. For example, during a car accident, another vehicle can capture an image of the incident and transmit it to the driver through a smart assistant, alerting other drivers and preventing further accidents. However, transmitting high-resolution images in real-time is not practical due to low latency requirements and limited bandwidth resources in vehicular networks. 

As a remedy, semantic communication technology offers a solution by enabling the extraction of essential information from images, substantially reducing the data transmitted and lowering communication delays \cite{9398576}. Moreover, transmitting this information helps protect sensitive details like faces and license plates in images. As illustrated in Fig.~\ref{Framework}(a), traditional generative AI-enabled vehicles can use semantic communication technology to extract information from images at the transmitter end and restore it using generative AI technology at the receiver end. This approach reduces transmitted data, decreasing communication delays and improving V2V communication efficiency in vehicular networks.

However, a challenge remains with existing generative AI models that rely only on essential text information: They suffer from uncertainty and unreliability, making them unable to provide precise road information to smart assistants in other vehicles. For instance, when a receiving vehicle receives a message: ``Two vehicles have been involved in a traffic accident near a green belt," even if a generative AI generates a high-quality image of the accident, there may be significant discrepancies between the generated image and the actual accident scene. This unreliability could result in vehicles receiving incorrect road information, which is unacceptable for safety-focused vehicular networks. Therefore, a more reliable and efficient generative AI framework is needed in vehicular networks to address these challenges and provide precise road information for better decision-making and enhanced safety.

\subsection{Multi-modality semantic-aware framework for generative AI-enabled vehicular networks}

In this section, we introduce a multi-modality semantic-aware framework for generative AI-enabled vehicular networks. This framework addresses the problem of large differences between transmitted and recovered images. Fig.~\ref{Framework}(b) shows the proposed framework, which uses both text data and image data to give more reliable guidance to receiving vehicles. By using the semantic skeleton of the transmitted image and text information, the generative AI-generated images of road conditions closely match the actual images.

{\bf Step 1. Semantic information extraction:}
Similar to the traditional generative AI-enabled framework, the first step is to extract semantic information from real-time road images. This process uses computer vision techniques to identify and classify objects in the images, such as vehicles, pedestrians, road signs, and traffic lights. Once the objects are identified, their attributes and characteristics, such as speed, direction, and location, are extracted to better understand the current road environment. Note that this extraction process can be optimized based on factors such as wireless transmission resources and the image reconstruction algorithms.

{\bf Step 2. Image skeleton extraction:}
Extracting the image skeleton is important because it provides a basic structure of the image and serves as the basis for extracting more advanced semantic information. This process involves identifying the edges and contours of the road condition image and creating a simplified representation of these features. The resulting image skeleton is a streamlined version of the original image that captures its essential features without unnecessary details. The image skeleton is then used for more advanced semantic analysis, including object recognition, lane detection, and traffic sign identification, making its extraction a key step in generating reliable and useful semantic information from road condition images.

{\bf Step 3. Wireless transmission: }
Upon extracting the image skeleton and text information, they are combined into a compact data package. This data package is then wirelessly transmitted to the receiving vehicle through V2V communication.  Note that the data volume of the skeleton information (i.e., approximately 0.5 megabytes) is relatively small compared with that of the original captured image (i.e., approximately 6.7 megabytes), and the less bandwidth is required for transmitting only the skeleton and text prompt. Consequently, the wireless transmission effectively ensures that the receiving vehicle acquires the fundamental information needed to reliably reconstruct the road condition image, while conserving network resources.

{\bf Step 4. Generative AI-enabled image generation: }
The generative AI model uses the extracted image skeleton and semantic information to generate a road condition image that closely resembles the actual image. The image skeleton provides the basic structure, while the semantic text information adds extra details such as road signs, vehicles, and pedestrians. By using both the image skeleton and semantic text information, the generative AI model can create a more reliable road condition image that helps vehicles make informed decisions on the road. Additionally, the model can generate images for various scenarios, such as different weather conditions, times of day, or traffic situations, giving a more complete understanding of road conditions. Note that this generation process can be optimized based on factors such as diffusion steps.

{\bf Step 5. Image reconstruction: }
The generated road condition image is reconstructed on the receiving side and provided to the intelligent assistant, which alerts the driver about potential hazards. The intelligent assistant analyzes the reconstructed image and compares it with the current road condition to identify potential hazards, such as accidents or roadblocks. If hazards are detected, the assistant alerts the driver through audio or visual cues, prompting them to take necessary precautions. This process allows the vehicle to stay informed about road conditions and take appropriate actions to ensure the safety of both its occupants and other vehicles on the road.

By following these steps, the proposed framework effectively reduces data transmission requirements and improves the reliability of road condition images in the vehicular networks. Additionally, the framework can enhance driving safety and reduce the occurrence of accidents, making it a valuable contribution to the field of vehicular networks.

\begin{figure*}[t!]
\centerline{\includegraphics[width=\textwidth]{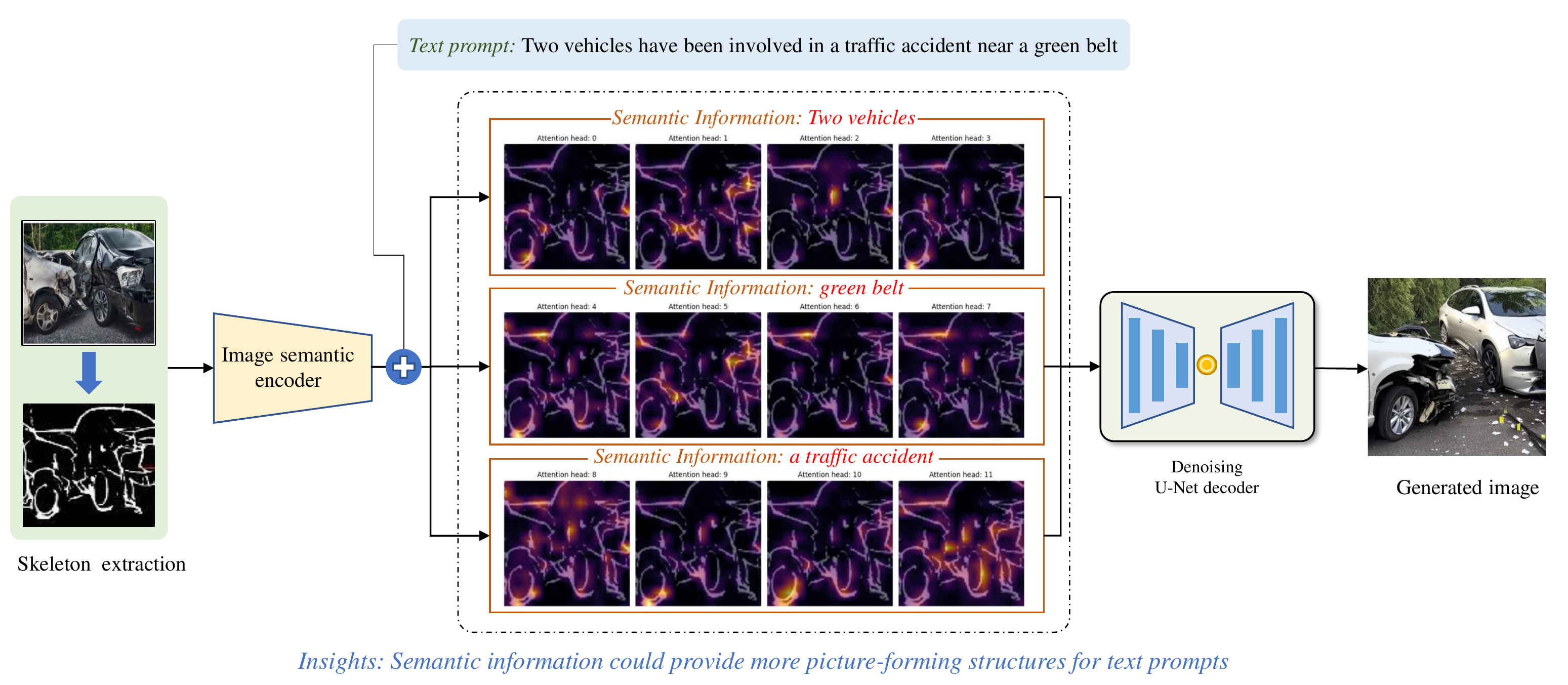}}
\caption{Compressing the image semantic information in multi-modality generative AI.}
\label{Semantic}
\end{figure*}
\section{Case study: Generative AI-enabled V2V resource allocation}
In this section, we propose a DRL-based approach to address the generative AI-enabled V2V resource allocation problem.

\subsection{System Model}
Following 3GPP V2X standards \cite{9345798}, we consider a generative AI-enabled vehicular networks consisting of multiple V2V links. The primary objective of this network is to ensure reliable and timely safety information sharing for each vehicle. To achieve this, orthogonal frequency division multiplexing technology is adopted and each V2V link has varying achievable transmission rates at different sub-channels.

In addition to the transmission rates, the image (i.e., skeleton) and text (i.e., prompt) information transmission frequencies also vary and dynamically change for each V2V link. We define the outage probability of successful transmission as a constraint, considering the advanced driving services of vehicle mobility. This constraint takes into account the achievable transmission rate, image similarity metric, channel coherence time, and generated image payload.

 Since transmission rate and image similarity are two critical indicators in generative AI-enabled V2V systems, we adopt a novel QoE metric based on the Weber-Fechner law to integrate these two indicators into the system optimization goal \cite{du2023attentionaware}. Specifically, the Weber-Fechner law describes the relationship between the magnitude of a physical stimulus and the perceived intensity, which in our case refers to the transmitted image quality. By incorporating this law into our QoE metric, we can account for both the transmission rate and the image similarity in a unified manner.  To optimize generative AI-enabled V2V resource allocation, we formulate the problem as follows: Our objective is to maximize the system QoE within the constraints of the transmission power budget and the probability of successful transmission for each vehicle. To achieve this, we need to optimize several parameters, including channel selection, transmission power for each vehicle, and the diffusion steps for inserting the skeleton. By doing so, we can effectively allocate resources to each V2V link in a way that maximizes system performance while ensuring reliable and regular safety information sharing among vehicles in the vehicular networks.


\subsection{Proposed DRL-based Approach and Simulations}

For the considered problem, we propose a double deep Q-network (DDQN)-based approach to address the resource allocation problem in generative AI-enabled V2V communication. DDQN is a deep reinforcement learning technique that employs two neural networks \cite{9575181}, namely the online network and the target network, to approximate the Q-value function. Specifically, the online network selects actions based on the current state, while the target network is utilized to estimate the Q-value of the subsequent state. The DDQN-based approach enhances the stability of the training process and prevents the overestimation of Q-values. Moreover, the action space, state space, and reward function are designed as follows.

\subsubsection{\bf Actions} The action space is composed of the  optimizing parameters, i.e., the selectable channel, the transmit power, and  the diffusion steps. Specifically, for the selectable sub-channel, binary variables indicate whether to select the sub-channel or not. For the transmit power and diffusion steps, feasible values are quantized into multiple values to accommodate practical circuit limitations and facilitate learning.

\begin{figure*}[t] 
	\centering  
	\subfigcapskip=1pt 
    \vspace{-0.5cm} 
	\subfigure[The obtained reward versus the number of iteration number with the different approaches.]{
		\label{Reward}
		\includegraphics[width=0.31\textwidth]{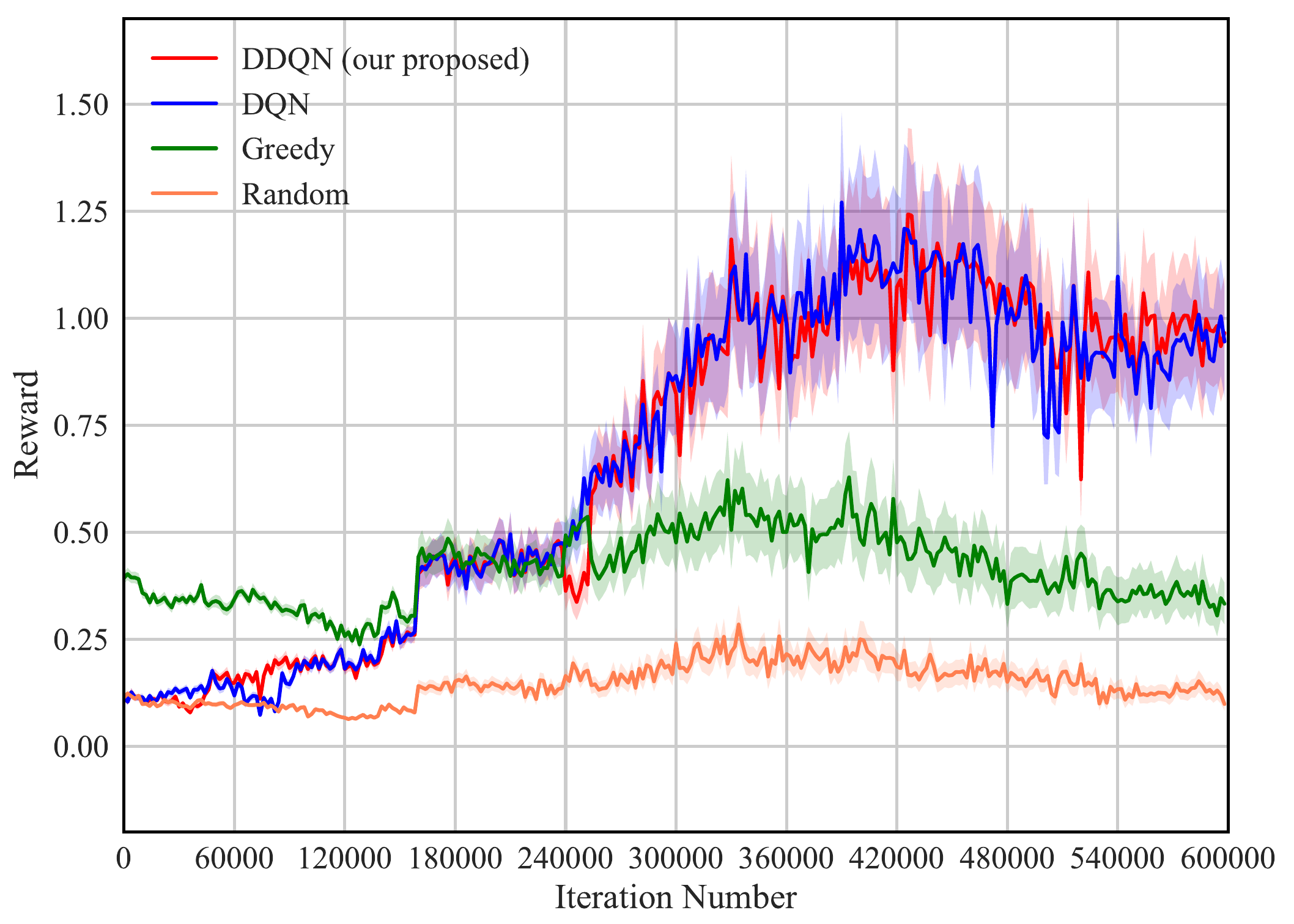}}
	\quad 
	\subfigure[The average QoE versus the size of image payload with the different approaches.]{
		\label{QoE}
		\includegraphics[width=0.30\textwidth]{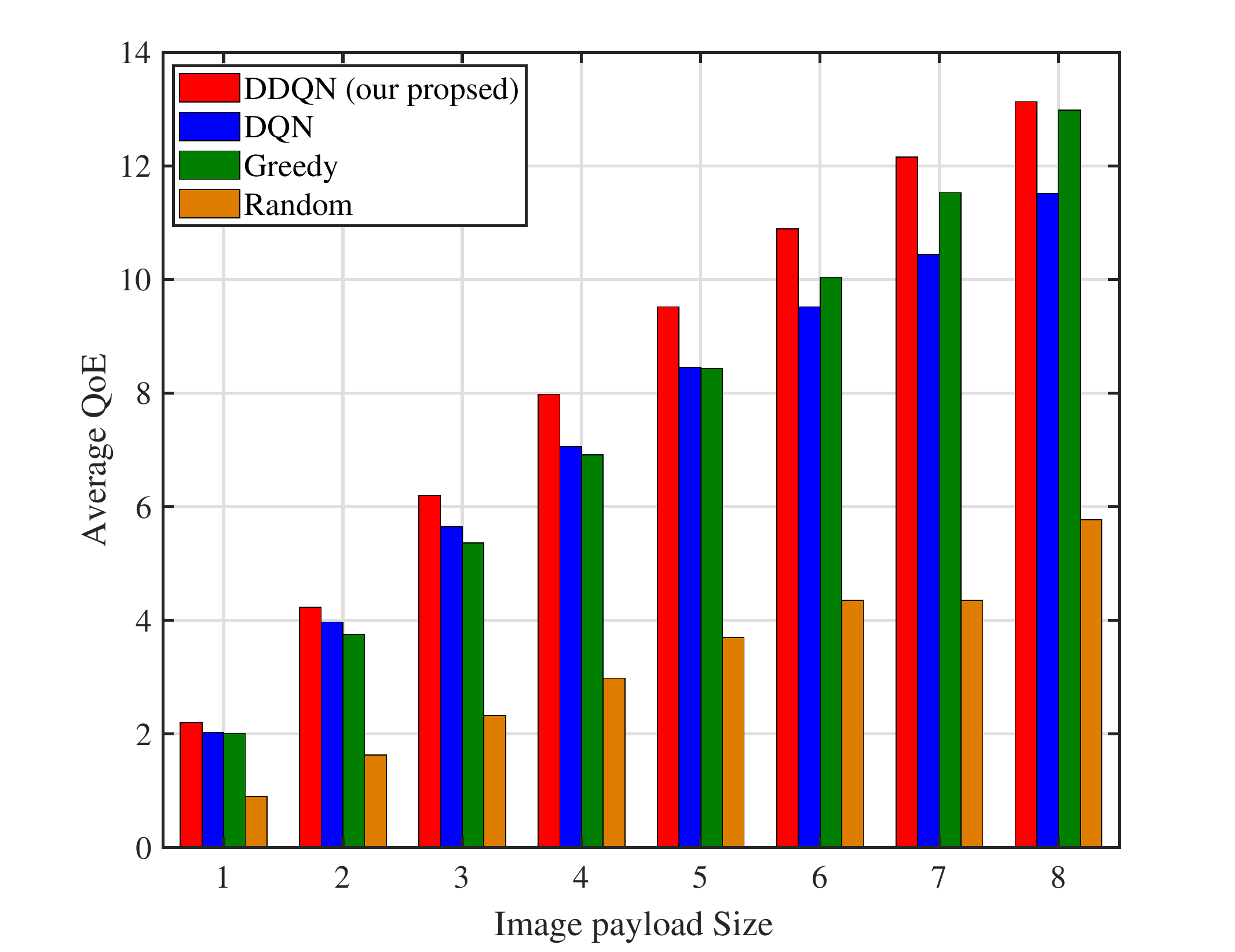}}
    \quad
    \subfigure[The average successful transmission data with the different approaches.]{
		\label{data}
		\includegraphics[width=0.30\textwidth]{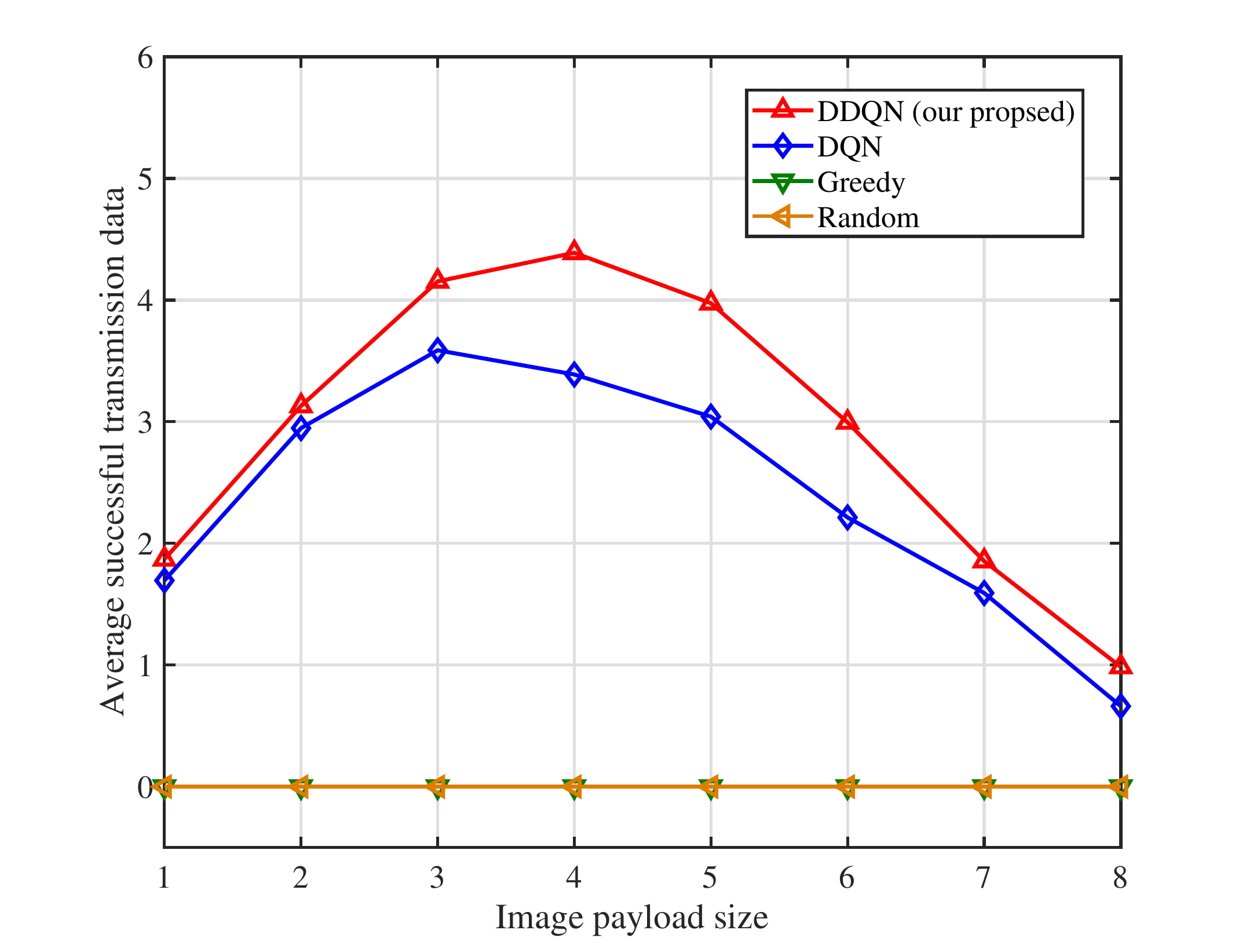}}
	\caption{Simulation results.}
	\label{fig RSMAandSDMA}
	\vspace{-0.5cm} 
\end{figure*}
\subsubsection{\bf States} In defining the state space, our aim is to incorporate as much relevant environmental information as possible for the considered problem \cite{10032267}. In the proposed DDQN-based approach, the state space consists of two parts: the current information and the previously selected action. Specifically, the current information depends on the channel information of each V2V link, the transmission rate of each V2V link, and the generated image payload. The previously selected action depends on the action taken during the previous time step.

\subsubsection{\bf Rewards} 
The reward function takes into account both the objective function and the constraints of the considered problem. Specifically, it consists of two terms. The first term is the instant reward, reflecting the unconstrained system QoE achieved at each time step. The second term is the penalty term, which assigns a larger penalty if the constraints are not satisfied. By combining these two terms, the reward function encourages the agent to achieve high QoE while simultaneously satisfying the constraints.

Fig.\ref{Semantic} shows an example of the multi-modality semantic-aware generative AI-enabled V2V, where the received image skeleton is extracted from the captured road condition image. It shows that in the received vehicles with the semantic encoder, the semantic information for the keywords of the prompt that generate the image is presented, where the image representation reflects the importance of each keyword through different brightness levels, with higher brightness indicating greater weight assigned to the corresponding keywords. The reason is that semantic information is designed to intelligently emphasize specific keywords (i.e., two vehicles, a green belt, and a traffic accident) during the image generation process, which facilitates the creation of more reliable traffic images based on the image skeletons and text prompts, leading to a clearer and more comprehensive understanding of the road conditions.


In the training stage, Fig.~\ref{Reward} compares of the average rewards obtained by four types of resource allocation strategies as the iteration number increases, where the corresponding curves are smoothed via a sliding window to provide a clearer overall trend of the raw data. It shows that the proposed DDQN-based approach consistently achieves higher rewards and converges more rapidly compared to the greedy-based and random-based approaches. Furthermore, although the rewards acquired by the proposed DDQN and DQN during the training phase are roughly similar, the reward value of the proposed DDQN surpasses that of DQN after convergence, and the fluctuations in the proposed DDQN-based approach are smaller than that in DQN. This observation can be attributed to the proposed DDQN-based approach's ability to prevent overestimation by decoupling the Q-target.



Next, in the testing stage, Fig.~\ref{QoE} shows the QoE achieved  by four types of resource allocation strategies as the size of image payload.  It is seen that the proposed DDQN-based approach is superior to the other benchmarks. It also shows that the increment of the size of image payload is beneficial to the system QoE, as a larger payload implies a more substantial amount of transmitted semantic information.

To further examine the impact of image payload size on received QoE, Fig.~\ref{data} is plotted. It is observed that the proposed DDQN-based approach is superior to other benchmarks, where the amount of data successfully transmitted under random and greedy are always 0. This is because the two methods do not guarantee that the successful transmission outage probability for each vehicle is satisfied. Moreover, it is interesting to find that as the image payload increases, the average amount of data for successful image transmission initially rises and then declines. The phenomenon may be attributed to vehicular networks congestion and elevated probability of packet loss (i.e., decreased transmission success rate) under high payload conditions.  Consequently, the number of retransmissions may increase, thereby reducing the overall transmission success rate.


\section{Future Directions}


\subsection{Integration with Edge and Fog Computing}
To address the challenges of resource constraints in generative AI-enabled vehicular networks, future research could explore the integration of edge and fog computing. By offloading some of the data processing and content generation tasks to nearby edge devices or fog nodes, vehicles can conserve resources while still benefiting from generative AI-based insights. This integration could lead to more efficient and reliable communication between vehicles and other connected devices.

\subsection{Sustainable and Energy-efficient Solutions}
As the transportation sector moves towards electrification and sustainability, generative AI-enabled vehicular networks can play a crucial role in promoting energy-efficient and environmentally friendly solutions. Future research could explore how generative AI can optimize routes, speeds, and driving behaviors to reduce energy consumption and emissions. Additionally, generative AI-enabled vehicular networks could facilitate the integration of electric vehicles (EVs) and renewable energy sources by intelligently managing charging schedules and energy consumption based on the availability of renewable energy and the demands of the grid. 

\subsection{Cross-domain Collaboration and Standardization}
As generative AI-enabled vehicular networks applications continue to expand, it will become increasingly important to establish collaboration and standardization across various domains. This includes the development of unified communication protocols and data formats to facilitate seamless interaction between vehicles, infrastructure, and other connected devices. Additionally, the collaboration between researchers, industry stakeholders, and policymakers will be essential to create comprehensive regulatory frameworks that ensure safety, privacy, and equitable access to generative AI-enabled vehicular networks services. 

\section{Conclusion}

In this article, we have introduced the generative AI-enabled vehicular networks, emphasizing the generative AI technologies and their application scenarios. Following this, we have proposed a multi-modality semantic-aware framework designed to enhance the service quality of generative AI, which encompassed both semantic and multi-modal technologies. To further improve system transmission efficiency, we have proposed a DRL-based approach to tackle the generative AI-enabled V2V resource allocation problem. Our simulation results have validated the effectiveness of the proposed approach. Finally, we have summarized potential research directions in the realm of generative AI-enabled vehicular networks.

\bibliographystyle{IEEEtran}
\bibliography{main}

\begin{thebibliography}{10}
\providecommand{\url}[1]{#1}
\csname url@samestyle\endcsname
\providecommand{\newblock}{\relax}
\providecommand{\bibinfo}[2]{#2}
\providecommand{\BIBentrySTDinterwordspacing}{\spaceskip=0pt\relax}
\providecommand{\BIBentryALTinterwordstretchfactor}{4}
\providecommand{\BIBentryALTinterwordspacing}{\spaceskip=\fontdimen2\font plus
\BIBentryALTinterwordstretchfactor\fontdimen3\font minus
  \fontdimen4\font\relax}
\providecommand{\BIBforeignlanguage}[2]{{%
\expandafter\ifx\csname l@#1\endcsname\relax
\typeout{** WARNING: IEEEtran.bst: No hyphenation pattern has been}%
\typeout{** loaded for the language `#1'. Using the pattern for}%
\typeout{** the default language instead.}%
\else
\language=\csname l@#1\endcsname
\fi
#2}}
\providecommand{\BIBdecl}{\relax}
\BIBdecl


\bibitem{8767077}
K.~Liu, X.~Xu, M.~Chen, B.~Liu, L.~Wu, and V.~C.~S. Lee, ``A hierarchical
  architecture for the future {I}nternet of {V}ehicles,'' \emph{IEEE Commun.
  Mag.}, vol.~57, no.~7, pp. 41--47, 2019.



\bibitem{Kim_2021_CVPR}
S.~W. Kim, J.~Philion, A.~Torralba, and S.~Fidler, ``Drive{GAN}: {T}owards a
  controllable high-quality neural simulation,'' in \emph{Proc. IEEE CVPR},
  June 2021.


\bibitem{Yang_2020_CVPR}
Z.~Yang, Y.~Chai, D.~Anguelov, Y.~Zhou, P.~Sun, D.~Erhan, S.~Rafferty, and
  H.~Kretzschmar, ``Surfelgan: {S}ynthesizing realistic sensor data for
  autonomous driving,'' in \emph{Proc. IEEE CVPR}, June 2020.


\bibitem{mou2023t2iadapter}
C.~Mou, X.~Wang, L.~Xie, Y.~Wu, J.~Zhang, Z.~Qi, Y.~Shan, and X.~Qie,
  ``{T2I}-{A}dapter: {L}earning adapters to dig out more controllable ability
  for text-to-image diffusion models,'' 2023.



\bibitem{8732370}
Y.~Jing, Y.~Yang, Z.~Feng, J.~Ye, Y.~Yu, and M.~Song, ``Neural style transfer:
  {A} review,'' \emph{IEEE Trans. Vis. Comput. Graph.}, vol.~26, no.~11, pp.
  3365--3385, 2020.


\bibitem{Kong_2018_CVPR}
S.~Kong and C.~C. Fowlkes, ``Recurrent pixel embedding for instance grouping,''
  in \emph{Proc. IEEE CVPR}, June 2018.



\bibitem{van2023chatgpt}
E.~A. van Dis, J.~Bollen, W.~Zuidema, R.~van Rooij, and C.~L. Bockting,
  ``Chat{GPT}: {F}ive priorities for research,'' \emph{Nature}, vol. 614, no.
  7947, pp. 224--226, 2023.



\bibitem{10.1145/2366145.2366154}
\BIBentryALTinterwordspacing
M.~Fisher, D.~Ritchie, M.~Savva, T.~Funkhouser, and P.~Hanrahan,
  ``Example-based synthesis of {3D} object arrangements,'' \emph{ACM Trans.
  Graph.}, vol.~31, no.~6, 2012. [Online]. Available:
  \url{https://doi.org/10.1145/2366145.2366154}
\BIBentrySTDinterwordspacing



\bibitem{pmlr-v80-jin18a}
\BIBentryALTinterwordspacing
W.~Jin, R.~Barzilay, and T.~Jaakkola, ``Junction tree variational autoencoder
  for molecular graph generation,'' in \emph{Proc. ICML}, 2018. [Online].
  Available: \url{https://proceedings.mlr.press/v80/jin18a.html}
\BIBentrySTDinterwordspacing



\bibitem{du2023enabling}
H.~Du, Z.~Li, D.~Niyato, J.~Kang, Z.~Xiong, Xuemin, Shen, and D.~I. Kim,
  ``Enabling AI-generated content {(AIGC)} services in wireless edge
  networks,'' 2023.


\bibitem{9398576}
H.~Xie, Z.~Qin, G.~Y. Li, and B.-H. Juang, ``Deep learning enabled semantic
  communication systems,'' \emph{IEEE Trans. Signal Process.}, vol.~69, pp.
  2663--2675, 2021.


\bibitem{9345798}
M.~H.~C. Garcia, A.~Molina-Galan, M.~Boban, J.~Gozalvez, B.~Coll-Perales,
  T.~Şahin, and A.~Kousaridas, ``A tutorial on {5G} {NR} {V2X}
  communications,'' \emph{IEEE Commun. Surveys Tuts.}, vol.~23, no.~3, pp.
  1972--2026, 2021.



\bibitem{du2023attentionaware}
H.~Du, J.~Liu, D.~Niyato, J.~Kang, Z.~Xiong, J.~Zhang, and D.~I. Kim,
  ``Attention-aware resource allocation and {QoE} analysis for metaverse
  x{URLLC} services,'' 2023.



\bibitem{9575181}
R.~Zhang, K.~Xiong, and Y.~Lu, \emph{et al.}, ``Joint coordinated beamforming
  and power splitting ratio optimization in {MU-MISO SWIPT}-enabled hetnets: A
  multi-agent {DDQN}-based approach,'' \emph{IEEE J. Sel. Areas Commun.},
  vol.~40, no.~2, pp. 677--693, 2022.




\bibitem{10032267}
R.~Zhang, K.~Xiong, Y.~Lu, P.~Fan, D.~W.~K. Ng, and K.~B. Letaief, ``Energy
  efficiency maximization in {RIS}-assisted {SWIPT} networks with {RSMA}: A
  {PPO}-based approach,'' \emph{{appear} to IEEE J. Sel. Areas Commun.}, 2023.


\end{thebibliography}
\end{document}